\pdfoutput=1
\documentclass[onecollarge,natbib]{svjour2}
\bibpunct{[}{]}{;}{n}{}{,} 
\smartqed  
\usepackage{graphicx}
\usepackage{hyperref}
\usepackage{float}
\usepackage{epsfig}
\usepackage{graphicx}
\usepackage{mathptmx}      
\usepackage{latexsym}
\usepackage{longtable}
\usepackage{dcolumn}
\usepackage{comment}
\usepackage{graphicx,color}

\usepackage{mathptmx}      
\journalname{Proceedings of Light Cone 2015}
\begin{document}

\title{Proton-Proton On Shell Optical Potential at High Energies and the Hollowness Effect
\thanks{Talk presented by ERA at the Light Cone 2015 Conference, Frascati, Italy. 21-25 September 2015.\\ 
Supported by Spanish DGI and Feder funds (grant FIS2014-59386-P), by Junta de Andaluc{\'{\i}a} (grant FQM225), and by Polish National Science Center (grant  DEC-2012/06/A/ST2/00390).     
}}


\author{ Enrique Ruiz Arriola   
       \and
        Wojciech Broniowski 
}


\institute{              Enrique Ruiz Arriola \at
              Departamento de F\'{\i}sica At\'{o}mica, Molecular y Nuclear and \\ Instituto Carlos I de  F{\'\i}sica Te\'orica y Computacional,
              Universidad de Granada, E-18071 Granada, Spain
           \and
Wojciech Broniowski \at
              The H. Niewodnicza\'nski Institute of Nuclear Physics, polish Academy of Sciences, PL-31342 Krak\'ow, Poland, and \\ 
              Jan Kochanowski University, PL-25406 Kielce, Poland \\
              \email{Wojciech.Broniowski@ifj.edu.pl} 
}

\date{\today}

\maketitle

\begin{abstract}
We analyze the usefulness of the optical potential as suggested by the
double spectral Mandelstam representation at very high energies, such
as in the proton-proton scattering at ISR and the LHC. Its particular
meaning regarding the interpretation of the scattering data up to the
maximum available measured energies is discussed. Our analysis
reconstructs 3D dynamics from the effective transverse 2D impact
parameter representation and suggests that besides the onset of gray
nucleons at the LHC there appears an inelasticity depletion
(hollowness) which precludes convolution models at the attometer
scale.
\end{abstract}


\section{Introduction \label{sec:intro}}

The history of proton-proton scattering at high energies has been
marked by continuous surprises; extrapolations have often been
contradicted by actual measurements such as at ISR (see,
e.g.,~\cite{Amaldi:1979kd,cheng1987expanding,Matthiae:1994uw,Barone:2002cv}
for comprehensive accounts and references therein). The first run of
the CERN-LHC on pp collisions has unveiled new features at CM energies
about $7$~TeV, measured by the TOTEM
collaboration~\cite{Antchev:2013gaa}. Contrary to naive expectations,
probing strong interactions at such high energies, corresponding to a
de Broglie wavelength of about half an attometer, becomes more
intricate and may still be far from the
asymptotics~\cite{Fagundes:2015lba}. Regge
phenomenolgy~\cite{Collins:1977jy} motivated Barger and
Phillips~\cite{Phillips:1974vt} to propose a parameterization which,
when suitably extended, describes all ISR and the LHC data
simultaneously~\cite{Fagundes:2013aja,Selyugin:2015pha}. Dynamical
calculations display the intricacy of non perturbative phenomena at
these high energies from a fundamental viewpoint (see
e.g. \cite{Khoze:2014aca,Gotsman:2014pwa}). In the present talk we
return to a phenomenological level and unveil interesting features of
pp scattering via the so-called {\em on-shell optical potential
model}.

The optical potential was first suggested to deal with inelastic
neutron-nucleus scattering above the compound nucleus
regime~\cite{Fernbach:1949zz}. There, the concept of the black disk
limit was first proposed and tested along with the observed Fraunhofer
diffraction pattern~\cite{blatt2012theoretical}, which also applies to
the eikonal approximation~\cite{glauber1959high}. The Serber
model~\cite{serber1964high} was an incipient extension of the optical
eikonal formalism to high energy particle physics. Based on a double
spectral representation of the Mandelstam representation of the
scattering amplitude, Cornwall and Ruderman delineated a definition of
the optical potential directly rooted in field
theory~\cite{cornwall1962mandelstam}. Field-theoretic discussions
using the multichannel Bethe-Salpeter equation shed some further light
~\cite{torgerson1966field,arnold1967optical} (see, e.g., an early
review on optical models~\cite{islam1972optical}). The off-shell vs
on-shell interplay was analyzed and an on-shell-type equation was
proposed by Namyslowski~\cite{namyslowski1967relativistic}. The high
energy grayness of the nucleon has been a matter of discussion since
the 70's~\cite{Kaidalov:1979jz}.

\section{Amplitudes and parameterizations \label{sec:intro}}

The invariant proton-proton elastic scattering differential cross section 
is given by 
\begin{eqnarray}
\frac{d\sigma_{\rm el}}{dt}= \frac{\pi}{p^2} \frac{d \sigma_{\rm el}}{d \Omega} = 
\frac{\pi}{p^2} |f(s,t) |^2 \, , 
\end{eqnarray}
where $f (s,t)$ is the scattering amplitude having both the partial
wave and the impact parameter
expansions~\cite{Blankenbecler:1962ez}
\begin{eqnarray}
f(s,t) =\sum_{l=0}^ \infty (2l+1) f_l(p) P_l(\cos \theta) 
=  \frac{p^2}{\pi} \int d^2 b \, h(\vec b,s) \, e^{i \vec q \cdot \vec b} = 
 2 p^2 \int_0^\infty b db J_0(bq) h(b,s) \, , 
\label{eq:PWA}
\end{eqnarray} 
where $s= 4(p^2 + M^2)$, $p$ is the CM momentum, and $t=-\vec q^2 $
with $q = 2p \sin (\theta/2) $ denoting the momentum transfer. In the eikonal
approximation one has $bp = l +1/2 + {\cal O}(s^{-1})$, hence
$h(b,s)=f_l(p) + {\cal O}(s^{-1})$ and $P_l(\cos \theta) \to J_0 (qb)$.
The total, elastic, and total inelastic cross sections
read, respectively~\cite{Blankenbecler:1962ez}
\begin{eqnarray} 
\sigma_T  &=& \frac{4 \pi}p {\rm Im} f(\theta=0) = 4 p \int d^2 b {\rm Im} h(\vec b,s) \, ,\\
\sigma_{\rm el} &=& \int d\Omega |f(s,t)|^2 = 4 p^2 \int d^2 b |h(\vec b,s)|^2 \, ,\\
\sigma_{\rm in} &\equiv& \sigma_T - \sigma_{\rm el} = \int d^2 b
n_{\rm in} (b)  \,.
\label{eq:sigmain}
\end{eqnarray} 
Here, the transverse probability inelastic profile
fulfills $n_{\rm in}(b)  \le 1$ and is given by
\begin{eqnarray}
n_{\rm in} (b)  = 4 p \left[ {\rm Im} h(b,s) - p |h(b,s)|^2 \right]  \, . 
\end{eqnarray} 
For our purposes we just need a working parameterization of the
scattering amplitude. Here, and for definiteness we use the work in
Ref.~\cite{Fagundes:2013aja}, and more specifically, their MBP2 form, 
which have been fitted
separately for all known differential cross sections for $\sqrt{s}=
23.4$, $30.5$, $44.6$, $52.8$, $62.0$, and $7000~{\rm GeV}$~\footnote{A
compilation of high energy scattering data can be found
at \url{http://www.theo.phys.ulg.ac.be/alldata-v2.zip}.} and read
\begin{eqnarray}
f(s,t) = \sum_{n} c_n F_n(t) s^{\alpha_n (t)} \, , \label{eq:BP}
\end{eqnarray}
where $F_n(t)$ are form factors, $\alpha_n (t)
= \alpha_n(0)+ \alpha_n'(0) t $ and $c_n$ are complex numbers which at
variance with Regge theory~\cite{Collins:1977jy}, are assumed to be
energy dependent. The fits produce $\chi^2/{\rm d.o.f.} \sim
1.2-1.7$~\cite{Fagundes:2013aja,Selyugin:2015pha}.

\section{The on-shell optical potential}

A general field theoretic approach requires solving a coupled
channel Bethe-Salpeter equation involving all open channels, a
most impractical a procedure, since their number 
becomes huge for the large energies at ISR or the LHC. Of course, a
viable approach would be to determine the kernel, operating as a
phenomenological optical potential, from the available NN scattering
data. In the geometric picture, the diffraction pattern is manifest as
a shadow of the inelastic scattering, such that the diffraction peak
in the forward direction is due to a coherent interference. From a
Quantum Mechanics point of view, the inelastic process can be
interpreted as a leakage in the probability current. We propose a
scheme below where a local and energy dependent potential can be
directly computed from the data, unveiling the structure of the {\em
inelasticity hole}.

A standard tool for handling the two-body relativistic scattering is the
Bethe-Salpeter equation (we use conventions of
Ref.~\cite{Nieves:1999bx}), which in the operator form reads
\begin{eqnarray}
T=V+V G_0 T \, . 
\label{eq:BS-op}
\end{eqnarray} 
For the $2 \to 2$ sector with the kinematics $(P/2+k,P/2-k) \to
(P/2+p,P/2-p) $ it can be written as
\begin{eqnarray}
T_P(p,k) = V_P(p,k) + {\rm i}\int\frac{d^4
q}{(2\pi)^4}T_P(q,k)S(q_+) S(q_-) V_P(p,q) \, ,
\label{eq:BS}
\end{eqnarray}
where $q_{\pm} = (P/2\pm q)$, $P$ is the total momentum, $T_P(p,k)$ is
the total scattering amplitude, and $S(q_\pm)$ denotes the nucleon propagator.
The kernel $V$ represents the irreducible four-point
Green's function, and it is generically referred to as the {\em potential}. Equation (\ref{eq:BS})
is a linear four-dimensional equation. It requires the
off-shell behavior of the potential $V_P(k',k)$ and generally
depends on the choice of the interpolating fields, although one expects
the scattering amplitude for the on-shell particles to be independent of
the field choice.\footnote{This point is not made very clear in the
literature; see, e.g., Ref.~\cite{Nieves:1999bx} for an explicit
demonstration in the particular case of $\pi\pi$ scattering that the field
transformations preserving the on-shell potential modify the on-shell
T-matrix.}

An approach which is manifestly independent of the off-shell
ambiguities deduces an on-shell equation by separating explicitly
those states which are on-shell and elastic from the
rest~\cite{namyslowski1967relativistic} (for a similar and related
ideas see, e.g., Ref.~\cite{Nieves:1999bx}). In the operator form, the
final result can be written as (see also
Ref.~\cite{cornwall1962mandelstam})
\begin{eqnarray}
T_{\rm el} = W + T_{\rm el} G_0 T_{\rm el}^\dagger \, ,
\label{eq:Nam}
\end{eqnarray}
where $W$ is the {\em on-shell optical potential} and $T_{\rm el}$ is the
elastic on-shell scattering amplitude, 
\begin{eqnarray}
T_{P,{\rm el}}(p,k) = T_{P}(p,k)|_{p^2=k^2=s/4-M^2} = 
- 8 \pi \sqrt{s} f(s,t) \, .
\end{eqnarray}
When written out explicitly, the equation becomes 
\begin{eqnarray}
T_{P,{\rm el}} (p,k) &=& W_P(p,k)
+{\rm i} \int\frac{d^4 q}{(2\pi)^4}
T_{P,{\rm el}} (q,k) S(q_+) S(q_-) \, T_{P,{\rm el}}(q,p)^* \, .
\end{eqnarray}
At the level of partial waves, we have a simplified form, which using
$p(s) = (s/4-M_N^2)^\frac12$ reads
\begin{eqnarray} 
f_l(s) &=& w_l(s) + \frac1{\pi} \int_{s_0}^\infty \, ds' \, \frac{f_l(s') p(s') f_l(s')^\dagger}{s'-s-i0^+}  \, ,
\end{eqnarray}  
where $s_0= 4M_N^2$. Note that only the on-shell amplitude enters
here, whereas the equation is non-linear. As usual, the scattering
amplitudes are defined as boundary values of analytic complex
functions, such that $f_l (s) \equiv f_l(s+i 0^+)$. Note that due to
the Schwartz reflection principle $f_l(s)^\dagger \equiv f_l(s-i 0^+)$,
and thus the unitarity condition corresponds to a right-hand
discontinuity cut $2 i {\rm Im} f_l(s) = f_l(s+i0^+) - f_l(s-i0⁺) $,
which reads
${\rm Im} \, f_l(s) - p(s) |f_l (s)|^2
= {\rm Im} w_l(s)$  at $s> 4 M_N^2$
and yields two contributions. The exchange, written as a left cut
condition becomes $ {\rm Im} f_l(s) = {\rm Im} w_l (s) $ for $ s< 0 $, 
whereas the causality implies a dispersion relation in energy along
the left and right cuts. Invoking the eikonal approximation, which
works phenomenologically for $\sqrt{s} \ge 23.5 {\rm GeV}$, and using
Eq.~(\ref{eq:sigmain}) we get
\begin{eqnarray}
 w_l(s)|_{l+1/2=bp} = n_{\rm in} (b)/4 p + {\cal O}(s^{-1})
\label{eq:wlb}
\end{eqnarray}

Solving the BS equation becomes very complicated and for a phenomenological
kernel is not truly essential.  Instead, we use a {\em minimal
relativistic approach}~\cite{Arriola:2014lxa} based on the squared mass
operator~\cite{Allen:2000xy} defined as
${\cal M}^2 = P^\mu P_\mu + {\cal V}$, where ${\cal V}$ represents the (invariant) interaction determined in
the CM frame by matching to the non-relativistic limit with a local
and energy-dependent phenomenological optical potential, $V(\vec r,s)=
{\rm Re } V(\vec r,s) + i {\rm Im} V(\vec r,s)$. This yields ${\cal
V}=8 M_N V(\vec r,s)$; it could be obtained by fitting
elastic scattering data~\cite{Arriola:2014lxa}.  After quantization we have 
$\hat{\cal  M}^2 = 4(\hat p^2+M_N^2) + 8 M_N V $, with 
$\hat p = -i \nabla$, such that the relativistic equation can be written as
$\hat{\cal  M}^2 \Psi = 4 (k^2+M_N^2) \Psi $, i.e., as a
non-relativistic Schr\"odinger equation
\begin{eqnarray}
(-\nabla^2 + M_N V ) \Psi = (s/4-M_N^2) \Psi \, . 
\end{eqnarray}
This equation incorporates the necessary physical ingredients which
were also present in the BS equation: relativity and inelasticity.
The optical potential $V$ {\it does not} yet correspond to the
on-shell one defined by Eq.~(\ref{eq:Nam}).  The argument given in
Ref.~\cite{cornwall1962mandelstam} uses the optical theorem from the
continuity equation, yielding
\begin{eqnarray}
\sigma_T - \sigma_{\rm el} \equiv \sigma_{\rm in}=- \frac{M_N}{p} \int d^3 x  {\rm Im} \, W( \vec x , s) \, ,
\label{eq:opti}
\end{eqnarray} 
where the on-shell optical potential is defined by its imaginary part
$ {\rm Im} \, W( \vec x , s) \equiv {\rm Im} \, V( \vec x , s)
|\Psi(\vec x)|^2 $ and can be interpreted
as the local density of inelasticity at a given CM energy $\sqrt{s}$;
it also becomes $W=V + \dots$ perturbatively. Using $\sum_{l=0}^\infty
(2l+1) \left[j_l(pr) \right]^2 = 1 $, Eq.~(\ref{eq:opti}) becomes
consistent with Eq.~(\ref{eq:wlb}). We may further rewrite this in
the impact parameter space by taking $\vec x=(\vec b,z)$ and integrating
over the longitudinal component. As a result we get Eq.~(\ref{eq:sigmain}),
where the  transverse probability profile function is given by 
\begin{eqnarray}
 n_{\rm in}(b) =-\frac{M_N}{p}\int_{-\infty}^\infty dz  \, {\rm Im} \, W(\vec x,s) =-\frac{M_N}{p}\int_{b}^\infty \frac{ 2 r dr}{\sqrt{r^2-b^2}} \, {\rm Im} \, W(r,s) \, .
\end{eqnarray} 
In last step the spherical symmetry has been exploited. To determine
$W(r,s)$ we recognize this formula as an integral equation of the Abel
type, hence it can be inverted using the standard method~(see
e.g. Ref.~\cite{buck1974inversion}) to give the on-shell optical
potential {\it directly} from the inelasticity profile and hence from
data,
\begin{eqnarray}
{\rm Im} W(r,s) =   \frac{2 p}{\pi M_N}\int_{r}^\infty \, db \, \frac{n_{\rm in}' (b)}{\sqrt{b^2-r^2}} \, . 
\end{eqnarray} 
This new formula is remarkable as it reconstructs the 3D on-shell
dynamics from the effective transverse 2D impact parameter
representation where the longitudinal physics has been integrated out.

\section{Numerical results and discussion}

\begin{figure}[h]
\begin{center}
\epsfig{figure=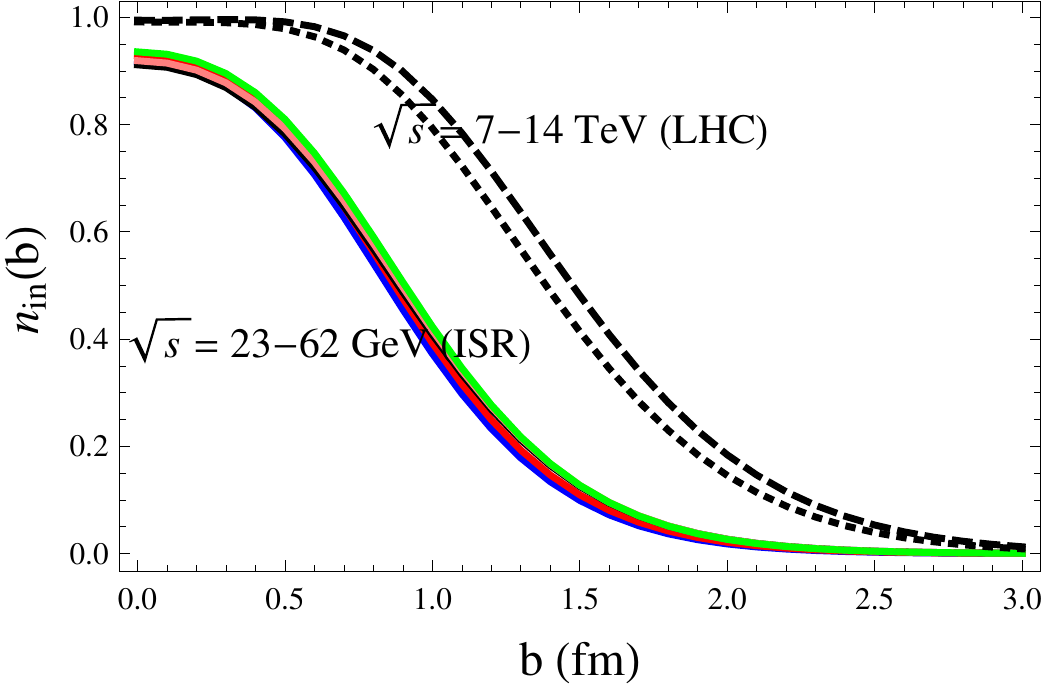,width=7cm,height=4.25cm}
\epsfig{figure=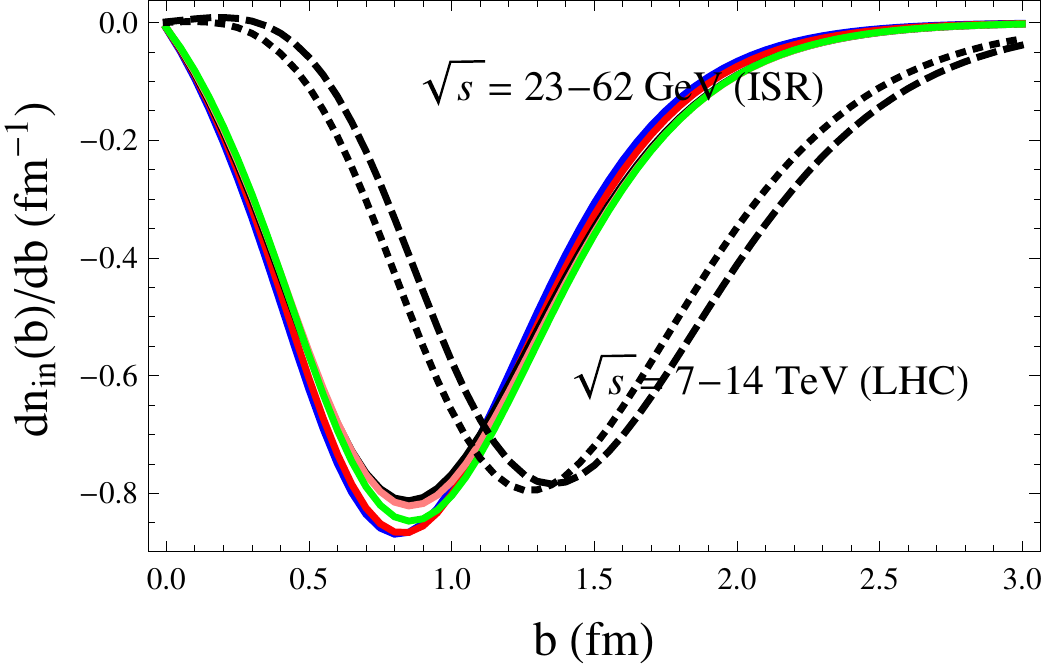,width=7cm,height=4.35cm}
\vskip.5cm
\epsfig{figure=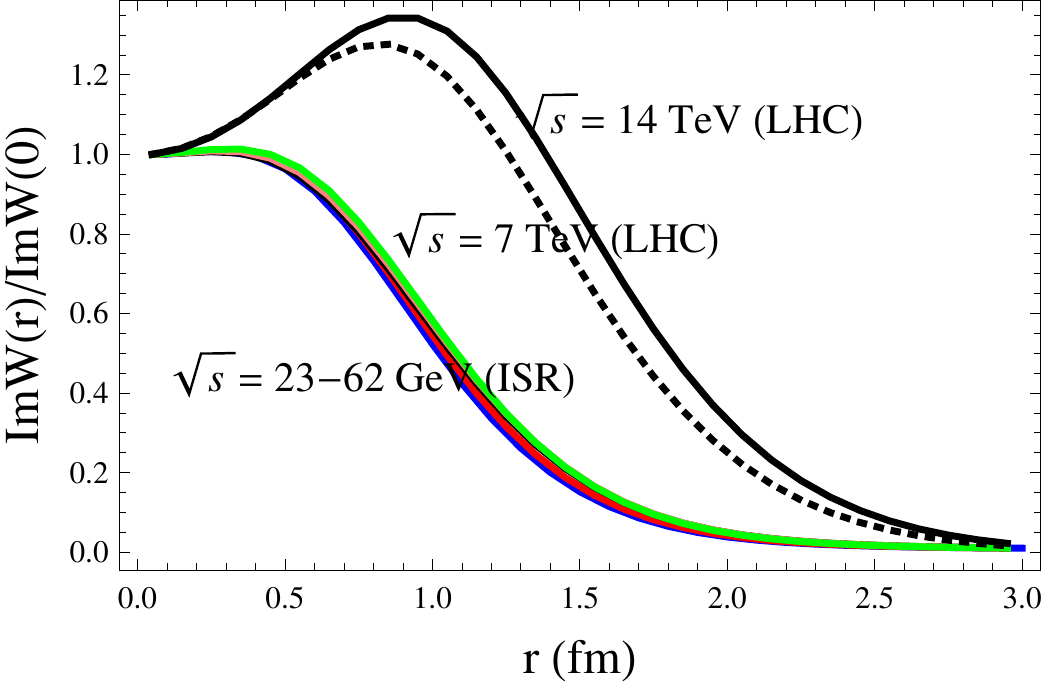,width=7cm,,height=4.35cm}
\end{center}
\caption{Inelasticity properties of the proton-proton scattering at the CM energies $\sqrt{s}=
23.4$, $30.5$, $44.6$, $52.8$, $62.0$, $7000$, and $14000~{\rm GeV}$.  Top: The inelastic profile (left) and its derivative (right) as a
function of the impact parameter. Bottom: The imaginary part of the
on-shell optical potential normalized to the value at the origin, plotted as a
function of the radial distance.}
\label{fig:ImW}
\end{figure}

For the MBP2 parameterization of Ref.~\cite{Fagundes:2013aja} we
obtain the inelastic profile function and its derivative for the
measured and fitted energies $\sqrt{s}= 23.4$, $30.5$, $44.6$, $52.8$,
$62.0$, and $7000~{\rm GeV}$.  The result for $14000~{\rm GeV}$ is an
extrapolation proposed in Ref.~\cite{Fagundes:2013aja}.  The
amplitudes of the on-shell potentials depend strongly on the CM energy
$\sqrt{s}$, with a power-like behavior. Thus, in Fig.~\ref{fig:ImW} we
show the ratio normalized to the value at the origin, ${\rm Im}
W(r,s)/{\rm Im} W(0,s)$. As can be vividly seen, the lower energy
values have a maximum at the origin, whereas the LHC pp data develop a
dip in the origin, which suggests that the inelasticity becomes
maximal at a finite value, around $r=1{\rm fm}$.  The fact that the
optical potential has its maximum away from the origin is most
remarkable; a feature shared by the profile function $n_{\rm in}(b)$,
which shows that in an inelastic collision most damage is not
necessarily produced by central collisions. The ``hollowness'' effect is
less evident in 2D as the 3D hole is integrated over the longitudinal
variables which effectively fill the 3D-hole.  This goes beyond the
idea that the protons become ``gray'' above $13 {\rm TeV}$, as
recently suggested by Dremin~\cite{Dremin:2015ujt} (see
also \cite{Dremin:2012ke,Troshin:2016svl}), rather than a black
disk. Actually, this shows in particular that the hollowness effect
cannot be reproduced by an intuitive folding structure. Indeed, for
small $r$ we get
\begin{eqnarray}
W(r) = \int d^3 y \rho( \vec y + \vec r/2) \rho(\vec y-\vec r/2) = \int d^3 y \rho(\vec y)^2 
- \frac14  \int d^3 y [\vec r \cdot \nabla \rho(\vec y)]^2 + \dots 
\end{eqnarray}
showing that $W(0)$ is a local maximum, in contrast to the
phenomenological result, see Fig.~\ref{fig:ImW}. This conclusion also
holds if the folding is made between wave functions with no extra
weight. Finally, we note that the mean squared radius of ${\rm
ImW(r)}$; $\langle r^2 \rangle =
\frac32 \langle b^2 \rangle$ displays a logarithmic growth with the energy.
These surprising new high energy features as well as the fluctuations
in $n_{\rm in}(b)$~\cite{Rybczynski:2013mla} become relevant in heavy
ions collisions and will be addressed in more detail elsewhere. The
energy interpolation of~\cite{Fagundes:2013aja} suggests that the 3D
depletion already happens at $0.5-1 {\rm TeV}$, below LHC energies,
generating a flattening of the 2D impact parameter dependence.

\section{Conclusions}

The on-shell optical potential is a meaningful concept under the most
common and general assumption of the Mandelstam double spectral
representation. We have shown that it is also a useful quantity when
interpreting the proton-proton scattering data at very high energies;
the shape of the inelasticity hole changes dramatically when going
from ISR to the LHC. The shoulder-like form of the imaginary part of
the on-shell potential resembles very much the traditional pattern
found in the absorptive part of optical potential in the
neutron-nucleus reactions beyond the compound model
regime~\cite{hodgson1971nuclear}. For a heavy nucleus, the surface is
much smaller than the volume and the shoulder merely shows that most
inelastic processes occur at the surface. This can pictorially be
imagined as derivatives of a Fermi-type distribution. In the case of
the proton-proton scattering, extremely high energies seem necessary
to resolve between the surface from the volume effects at the
attometer scales. The puzzling hollowness effect sets in at $0.5-1
{\rm TeV}$ and awaits a dynamical explanation.


\end{document}